\pdfoutput=1
\documentclass[11pt]{article}

\usepackage[final]{acl}
\usepackage{times}
\usepackage{latexsym}
\usepackage{multirow} 
\usepackage{array}
\usepackage{makecell}
\usepackage{subcaption}
\usepackage{enumitem}

\usepackage{pifont}
\usepackage{threeparttable}
\usepackage{adjustbox}
\usepackage{amsmath}
\usepackage[T1]{fontenc}
\usepackage[utf8]{inputenc}
\usepackage{booktabs} 
\usepackage{float}
\usepackage{microtype}
\usepackage[table]{xcolor} 
\newcommand{\strip}{\rowcolor{gray!8}}   % handy shortcut
\usepackage{inconsolata}

\usepackage{graphicx}
\definecolor{gain}{RGB}{0,128,0}  % dark green for improvements

\newcommand{\perf}[2]{\makecell{#1\\[-2pt]\textcolor{gain}{#2}}}

\renewcommand{\arraystretch}{0.95}
\title{\textsc{LLMInit}: A Free Lunch from Large Language Models \\ for Selective Initialization of Recommendation}

\author{
 \textbf{Weizhi Zhang\textsuperscript{1}},
 \textbf{Liangwei Yang\textsuperscript{2}\thanks{Corresponding author: liangwei.yang@salesforce.com}},
 \textbf{Wooseong Yang\textsuperscript{1}},
 \textbf{Henry Peng Zou\textsuperscript{1}},
 \\
 \textbf{Yuqing Liu\textsuperscript{1}},
 \textbf{Ke Xu\textsuperscript{1}},
 \textbf{Sourav Medya \textsuperscript{1}},
 \textbf{Philip S. Yu\textsuperscript{1}},
\\
\\
 \textsuperscript{1}University of Illinois Chicago,
 \textsuperscript{2}Salesforce AI Research
}

\begin{document}
\maketitle
\begin{abstract}

Collaborative filtering (CF) is widely adopted in industrial recommender systems (RecSys) for modeling user-item interactions across numerous applications, but often struggles with cold-start and data-sparse scenarios. Recent advancements in pre-trained large language models (LLMs) with rich semantic knowledge, offer promising solutions to these challenges. However, deploying LLMs at scale is hindered by their significant computational demands and latency. 
In this paper, we propose a novel and scalable LLM-RecSys framework, LLMInit, designed to integrate pretrained LLM embeddings into CF models through selective initialization strategies. Specifically, we identify the embedding collapse issue observed when CF models scale and match the large embedding sizes in LLMs and avoid the problem by introducing efficient sampling methods, including, random, uniform, and variance-based selections. Comprehensive experiments conducted on multiple real-world datasets demonstrate that LLMInit significantly improves recommendation performance while maintaining low computational costs, offering a practical and scalable solution for industrial applications. To facilitate industry adoption and promote future research, we provide open-source access to our implementation at \textcolor{blue}{\url{https://github.com/DavidZWZ/LLMInit}}.
\end{abstract}

\section{Introduction}

\begin{figure}[htpb]
    \centering
    \includegraphics[width=0.8\linewidth]{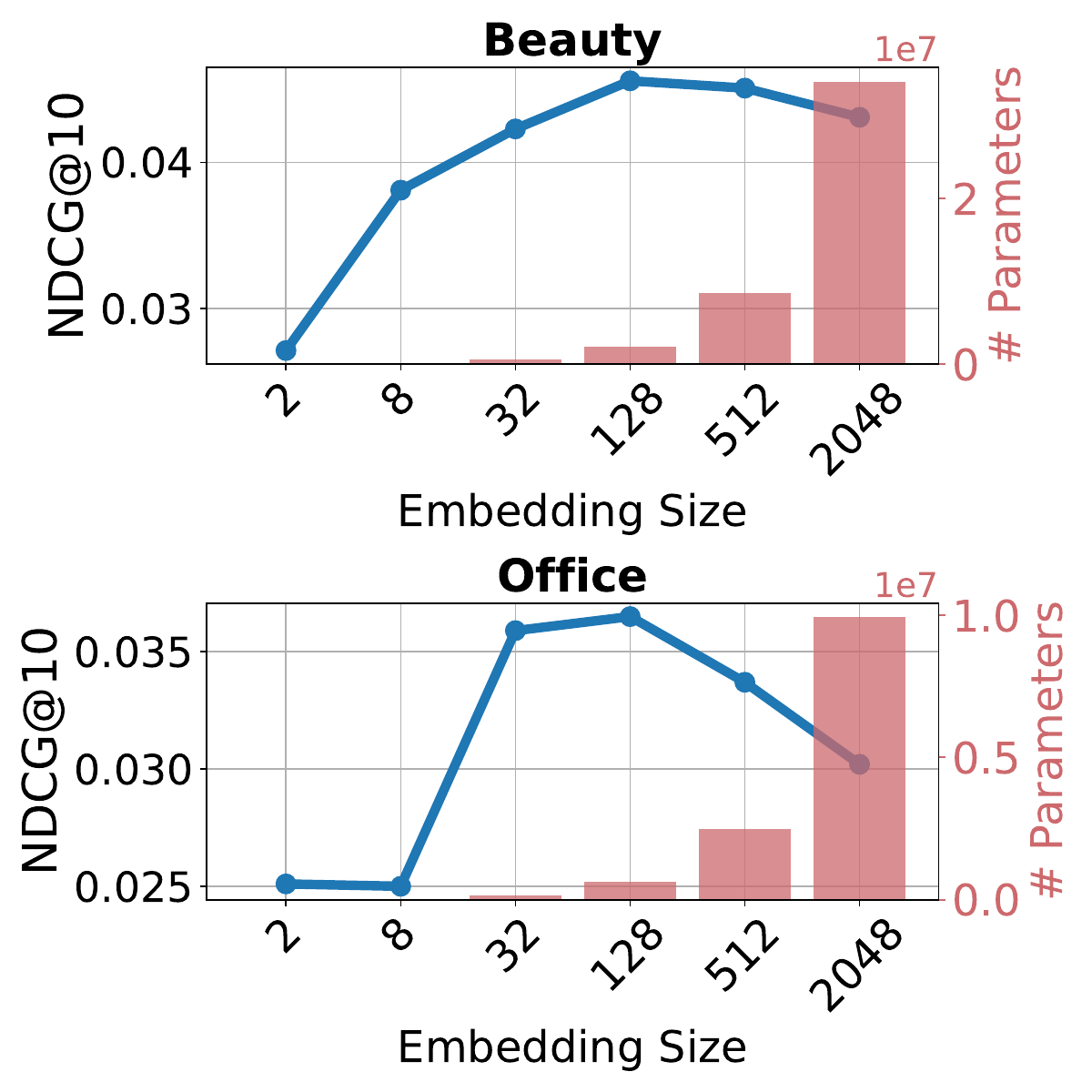}
    \caption{An investigation of the embedding collapse issue of CF model LightGCN~\cite{he2020lightgcn} in two Amazon-review datasets. Increasing embedding size with exponentially growing parameters finally leads to performance degradation.
    }
    \label{fig: emb_size}
    \vspace{-5pt}
\end{figure}

Recommender systems (RecSys), particularly collaborative filtering (CF), play a pivotal role in many online platforms and web applications ~\cite{koren2021advances, fan2019graph}, aiding users in navigating vast information by offering personalized suggestions. Among the various techniques, collaborative filtering~\cite{wang2019neural, he2020lightgcn, zhang2024we} stands out for its effectiveness in predicting user preferences by analyzing past user-item interactions. A key principle among these models is learning randomly initialized user/item representations with carefully designed loss functions to capture each user's preference \cite{rendle2012bpr}. However, such CF approaches heavily rely on the observed user-item interactions to optimize the CF embeddings from random initialization and perform poorly in real-world applications where interaction data is sparse and users/items are cold-start.

More recently, the rapid advancement of large language models (LLMs) such as GPT-4 ~\cite{achiam2023gpt} and LLaMa ~\cite{touvron2023llama}, which have exhibited remarkable proficiency in understanding and processing textual information, have piqued the interest of researchers seeking to explore their potential in improving recommender systems beyond traditional approaches ~\cite{geng2022recommendation, li2023text, zhang2025personaagent}. One of the most promising directions in this exploration involves adapting LLMs as recommender systems through prompt engineering and tailored instruction tuning. Such integration of LLMs enables the RecSys to better understand user preferences, and interpret contextual information, thereby surpassing the traditional models in cold-start settings.

Despite the inspiring progress made in LLMs for recommendations, their adoption in online deployments faces significant challenges related to efficiency and scalability. These issues stem from the inherently time-consuming and computationally intensive nature of LLMs. Such problems can be further exaggerated in real-world applications of large-scale users and items due to the enlarged vocabulary tokens and in-context input. Moreover, LLMs fall short of complex CF interaction understanding owing to the next-token prediction pipeline and the constraints of a maximum token limit (e.g., 4096 tokens in LLaMA-2) to incorporate the whole CF interactions. This limitation hampers their ability to capture and effectively model global user-item dependencies as in CF models, leading to poor performance in the full-ranking \cite{hou2024large} and warm-start settings \cite{bao2023tallrec}. 
Therefore, we raise a critical research question: 
\begin{center}
\textit{How can we harness the power of LLMs for RecSys in an effective and scalable manner?}
\end{center}

To answer the above question, we propose to utilize LLMs for CF-based RecSys from a new perspective by taking pretrained large language models as a free lunch for the embedding initialization of CF models. 
One trivial way is to directly utilize the LLMs to generate embeddings with substantial textual information. However, storing the large-scale embeddings from LLMs for each user and item is not efficient nor scalable, especially when the platform scales to millions of users and items. 
Furthermore, as illustrated in Figure~\ref{fig: emb_size}, unlike the scaling laws observed in language models \cite{kaplan2020scaling} and graph models~\cite{liu2024towards}, the performance of the recommendation model LightGCN~\cite{he2020lightgcn} deteriorates even when the numbers of parameters scale up exponentially. This behavior resembles the embedding collapse issue \cite{guo2024embedding} observed in click-through rate prediction models \cite{wang2021dcn}. These findings suggest that significantly smaller embedding sizes (e.g., 128) are more suitable for recommendation tasks. In contrast, the top-10 models in the LLM embedding benchmark MTEB~\cite{muennighoff2023mteb}, pretrained for various NLP tasks, use an average large embedding size of 4,506.

\begin{figure}[htpb]
    \centering
    \includegraphics[width=1\linewidth]{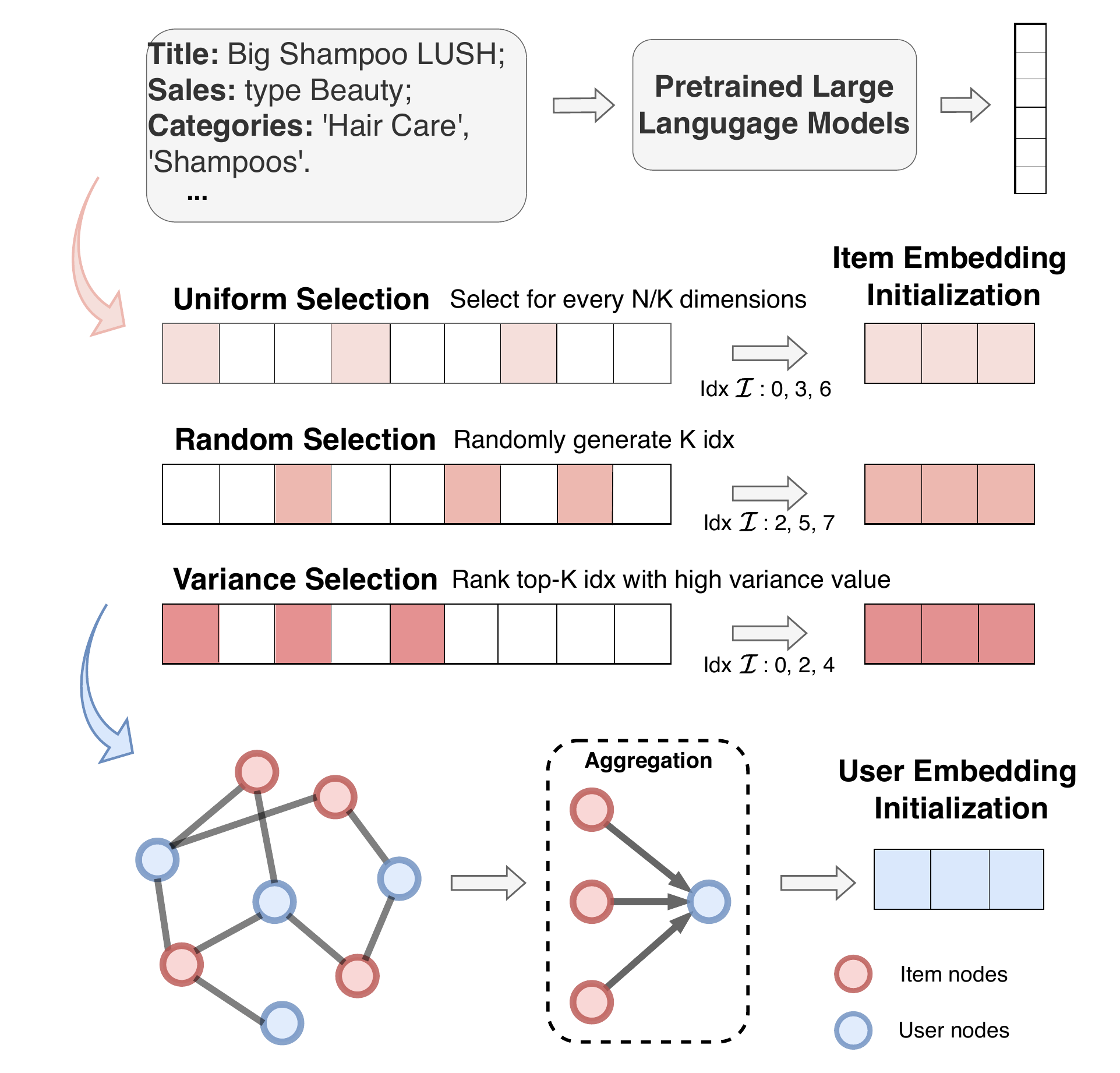}
    \caption{An Illustration of LLMInit framework including contextual LLM input, three types of selective item embedding initialization strategies, and the user embedding aggregation operation. 
    }
    \label{fig: method}
\end{figure}

In this paper, we present LLMInit, a practical and scalable initialization framework that bridges large language model (LLM) embeddings with collaborative filtering (CF) models for industrial-scale recommendation systems. LLMInit tackles the deployment challenges of LLMs and enhances the semantic capacity of CF models by initializing CF embedding models based on pretrained LLMs.
To this end, we introduce three efficient strategies, including random, uniform, and variance-based index selection, that selectively sample and compress LLM-generated item embeddings to suit the constrained embedding spaces of CF models. On the user side, where contextual signals may be missing or sparse, LLMInit aggregates item-level embeddings using lightweight pooling mechanisms. This design enables recommendation models to inherit rich language semantics from pretrained LLMs while maintaining high efficiency in both training and inference.
Extensive evaluations on multiple real-world datasets demonstrate that LLMInit consistently enhances performance across several state-of-the-art CF architectures. Compared to full LLM-based recommendation systems, our framework achieves significant gains in scalability and efficiency, making it well-suited for real-world deployments. Our contributions underscore the value of adapting language model innovations to meet the practical demands of large-scale, industry-grade recommendation systems.

\begin{itemize}[leftmargin=*]
\item Conceptually, we identify the embedding collapse issue in the CF-based recommendations and propose a new paradigm of effectively leveraging LLMs for CF embedding initialization.
\item Methodologically, we devise three types of selective initialization strategies to inherit the rich pretrained knowledge from LLMs into lightweight CF models for scalable recommendations.
\item Empirically, we implement LLMInit in a plug-and-play manner on various SOTA CF models, achieving significant performance and efficiency gains, especially compared to LLM RecSys.
\end{itemize}

\vspace{8pt}

\section{Preliminaries}

\subsection{Problem Definition}
Recommender systems (RecSys) aims to improve the platform's accuracy in personalization by leveraging content features and user-item interactions. Formally, consider a textual bipartite graph $\mathcal{G} = (\mathcal{V}, \mathcal{E})$ that models user-item interactions with node-level content features. The node set $\mathcal{V}$ is divided into users $u_i \in \mathcal{V}_u$ and items $v_j \in \mathcal{V}_v$. Edges $(u_i, v_j) \in \mathcal{E}$ represent interactions or relationships between users and items. The objective is to develop a recommendation algorithm that utilizes the node features and graph structure to predict and rank items that a user is likely to be interested in but has not yet interacted with.

% \vspace{-10pt}

\subsection{Related Work}

\textbf{LLMs for Recommendation.} 
Recent advancements in this domain can be categorized into three paradigms~\cite{wu2024survey}: extracting embeddings for sequential recommendation~\cite{qiu2021u,hou2022towards}, generating semantic tokens to capture user preferences~\cite{li2023text,xi2024towards}, and directly adapting LLMs as RecSys through prompting~\cite{hou2024large,liang-etal-2025-taxonomy} or tuning~\cite{geng2022recommendation, yang2024item}.  
However, they all fail to capture the intricate CF relationship in the user-item bipartite graphs as they either only focus on sequential order patterns \cite{qiu2021u,hou2022towards} or semantic understandings \cite{li2023text,xi2024towards, zhang2023dual}. In addition, 
they face notable challenges in efficiency and scalability due to the computational demands of LLMs, large-scale embeddings, overloaded vocabulary tokens, and limited in-context inputs in real-world scenarios. 
In contrast, our work LLMInit, proposes a novel paradigm to leverage LLMs in a scalable manner and addresses ineffective interaction modeling by resorting to CF models.

\begin{table*}[htbp]
  % \small
  \caption{Performance comparison on four datasets.  Gains are relative to the base model’s performance.}
  \label{tab:main}
  \centering

  % start alternating grey rows after the second line ( = first “Base” row )
  % \rowcolors{2}{gray!8}{white}

  \begin{adjustbox}{max width=\textwidth}
  \begin{tabular}{llcccccccc}
    \toprule
    \multirow{2}{*}{Method}
      & \multicolumn{2}{c}{Beauty}
      & \multicolumn{2}{c}{Toys–Games}
      & \multicolumn{2}{c}{Tools–Home}
      & \multicolumn{2}{c}{Office–Products}\\
    \cmidrule(lr){2-3}\cmidrule(lr){4-5}
    \cmidrule(lr){6-7}\cmidrule(lr){8-9}
      & R@10 & N@10 & R@10 & N@10 & R@10 & N@10 & R@10 & N@10\\
    \midrule
% ----------------------- LightGCN -----------------------
% \shortstack[l]{LightGCN\\\citep{he2020lightgcn}}

  LightGCN~\citep{he2020lightgcn}            & 0.0910 & 0.0432 & 0.0775 & 0.0360 & 0.0574 & 0.0283 & 0.0745 & 0.0365 \\
  \strip
  +LLMInit–Rand   & \perf{0.0960}{+5.5\%} & \perf{0.0467}{+8.1\%}
                    & \perf{0.0805}{+3.9\%} & \perf{0.0387}{+7.5\%}
                    & \perf{0.0612}{+6.6\%} & \perf{0.0313}{+10.6\%}
                    & \perf{0.0773}{+3.8\%} & \perf{0.0387}{+6.0\%} \\
  +LLMInit–Uni    & \perf{0.1006}{+10.6\%} & \perf{0.0469}{+8.6\%}
                    & \perf{0.0806}{+4.0\%}  & \perf{0.0388}{+7.8\%}
                    & \perf{0.0633}{+10.3\%} & \perf{\textbf{0.0319}}{+12.7\%}
                    & \perf{0.0791}{+6.2\%} & \perf{0.0395}{+8.2\%} \\
\strip
  +LLMInit–Var    & \perf{\textbf{0.1019}}{+12.0\%} & \perf{\textbf{0.0485}}{+12.3\%}
                    & \perf{\textbf{0.0808}}{+4.3\%}  & \perf{\textbf{0.0389}}{+8.1\%}
                    & \perf{\textbf{0.0633}}{+10.3\%} & \perf{0.0317}{+12.0\%}
                    & \perf{\textbf{0.0816}}{+9.5\%} & \perf{\textbf{0.0414}}{+13.4\%} \\
\midrule
% ----------------------- SGL -----------------------
{SGL~\cite{wu2021self}}      & 0.1017 & 0.0474 & 0.0832 & 0.0380 & 0.0580 & 0.0284 & 0.0669 & 0.0297 \\
\strip
  +LLMInit–Rand   & \perf{0.1069}{+5.1\%} & \perf{0.0520}{+9.7\%}
                    & \perf{0.0885}{+6.4\%} & \perf{0.0418}{+10.0\%}
                    & \perf{\textbf{0.0692}}{+19.3\%} & \perf{0.0337}{+18.7\%}
                    & \perf{\textbf{0.0810}}{+21.1\%} & \perf{\textbf{0.0426}}{+43.4\%}\\
  +LLMInit–Uni    & \perf{0.1101}{+8.3\%} & \perf{0.0513}{+8.2\%}
                    & \perf{0.0920}{+10.6\%} & \perf{0.0424}{+11.6\%}
                    & \perf{0.0676}{+16.6\%} & \perf{0.0333}{+17.3\%}
                    & \perf{0.0773}{+15.6\%} & \perf{0.0350}{+17.9\%}\\
\strip
  +LLMInit–Var    & \perf{\textbf{0.1106}}{+8.8\%} & \perf{\textbf{0.0530}}{+11.8\%}
                    & \perf{\textbf{0.0927}}{+11.4\%} & \perf{\textbf{0.0427}}{+12.4\%}
                    & \perf{0.0686}{+18.3\%} & \perf{\textbf{0.0339}}{+19.4\%}
                    & \perf{0.0794}{+18.7\%} & \perf{0.0421}{+41.8\%}\\
\midrule
% ----------------------- SGCL -----------------------
SGCL~\cite{zhang2025sgcl}           & 0.1027 & 0.0499 & 0.0828 & 0.0382 & 0.0585 & 0.0294 & 0.0647 & 0.0298 \\
\strip
  +LLMInit–Rand   & \perf{0.1094}{+6.5\%} & \perf{0.0512}{+2.6\%}
                    & \perf{0.0929}{+12.2\%} & \perf{0.0418}{+9.4\%}
                    & \perf{\textbf{0.0651}}{+11.3\%} & \perf{0.0326}{+10.9\%}
                    & \perf{0.0770}{+19.0\%} & \perf{0.0365}{+22.5\%}\\
  +LLMInit–Uni    & \perf{\textbf{0.1115}}{+8.6\%} & \perf{0.0513}{+2.8\%}
                    & \perf{0.0923}{+11.5\%} & \perf{\textbf{0.0422}}{+10.5\%}
                    & \perf{0.0650}{+11.1\%} & \perf{0.0327}{+11.2\%}
                    & \perf{0.0742}{+14.7\%} & \perf{0.0354}{+18.8\%}\\
\strip
  +LLMInit–Var    & \perf{0.1104}{+7.5\%} & \perf{\textbf{0.0522}}{+4.6\%}
                    & \perf{\textbf{0.0941}}{+13.6\%} & \perf{0.0421}{+10.2\%}
                    & \perf{0.0646}{+10.4\%} & \perf{\textbf{0.0327}}{+11.2\%}
                    & \perf{\textbf{0.0776}}{+19.9\%} & \perf{\textbf{0.0366}}{+22.8\%}\\
    \bottomrule
  \end{tabular}
  \end{adjustbox}
\end{table*}

\section{Method}
As in Figure~\ref{fig: method} LLMInit, before CF model training, the raw metadata is concatenated and fed to the LLMs. Then the semantic latent embeddings are selectively sampled via one of the CF item embedding initialization approaches followed by user embedding aggregation.

\subsection{Selective Item Embedding Initialization}
The CF item representation space can be regarded as a subspace of the sophisticated world-knowledge representation space in large language models \cite{sheng2024language}. Motivated by this, we propose to distill and selectively utilize the embeddings from LLMs to initialize the embeddings for CF recommendation models.

Formally, $ \mathbf{v} \in \mathbf{R}^N $ represents the item embedding vector generated by the pre-trained LLMs.
Let $ K $ denotes the desired embedding dimensionality after selection.
Then $ \mathcal{I} \subseteq \{0, 1, \dots, N-1\} $ is the selected indices.
The resulting $ K $-dimensional embedding is:
$
\mathbf{v}_\text{selected} = \{ \mathbf{v}[i] : i \in \mathcal{I} \}.
$
Here, we propose three novel initialization strategies to mitigate the inefficiency and embedding collapse issue for directly adopting the LLMs embeddings as follows.

\subsubsection{Uniform Selection (LLMInit-Uni)}
The uniform space among selections provides a balanced representation and helps mitigate potential redundancy or over-reliance on the LLM embeddings. By evenly spacing the indices, the strategy ensures that no region of the embedding vector dominates or is neglected across $ \mathbf{v} $:
$$
\mathcal{I} = \{ k \cdot \lfloor N / K \rfloor : k \in \{0, 1, \dots, K-1\} \}.
$$

\subsubsection{Random Selection (LLMInit-Rand)}
Randomization introduces diversity in the selected dimension patterns.
Indices $K$ are randomly sampled from $ \{0, 1, \dots, N-1\} $ based on the uniform distribution:
$$
\mathcal{I} \sim \text{Random}(\{0, 1, \dots, N-1\}), \quad |\mathcal{I}| = K.
$$

\subsubsection{Variance Selection (LLMInit-Var)}
Variance can serve as a heuristic for identifying information-rich and discriminative dimensions in the representation space. By selecting the top-$k$ dimensions of $\mathbf{v}$ with the highest variance across the dataset, this method prioritizes the most distinctive features, improving the separation of potential positive items from candidate pools.
Let $ \sigma^2[j] $ denote the variance of the $ j $-th dimension of $ \mathbf{h} $ across the dataset. The variance selects the embedding indices as: 
$$
\mathcal{I} = \text{Top-}K(\sigma^2[0], \sigma^2[1], \dots, \sigma^2[N-1]).
$$

\subsection{Aggregated User Embedding Initialization}

Based on the item-side embedding, we design an aggregation strategy for user embedding initialization to tackle the privacy situation where the user's context information is missing.
Suppose we observe a user with historical items initial embeddings $\mathbf{v}_j$, and we design the aggregation initialization based on the smoothed neighborhood pooling process. The user embedding $\mathbf{u}_i$ is finalized as:
\begin{equation}\label{eq: discrete}
\begin{aligned}
\mathbf{u}_i = \sum_{j\in N_i} \frac{1}{{|N_i|}}\mathbf{v}_j ,\\
% \mathbf{v}^{k+1}_j = \sum_{i\in N_j} \frac{1}{\sqrt{|N_j|} \sqrt{|N_i|}}\mathbf{u}^{k}_i,\\
\end{aligned}
\end{equation}
The normalization employs the degree ${|N_i|}$ to temper the magnitude and bias towards popular users after aggregation pooling.

\section{Experiments}
\subsection{Experimental Set Up}

We adopt the 5-core setting from \cite{zhang2025sgcl, zhang2024mixed} and conduct experiments on four real-world Amazon datasets~\cite{ni2019justifying}, 
\footnote{\url{https://jmcauley.ucsd.edu/data/amazon/links.html}},
including Beauty, Toys-Games, Tools-Home, and Office-Products, as summarized in Table~\ref{tab:data}. We use a leave-one-out strategy for data splitting and evaluate all the models using top-K metrics for ranking, including Recall@10 and NDCG@10.
For collaborative filtering (CF) baselines, we include LightGCN~\cite{he2020lightgcn} as a representative graph-based model, SGL~\cite{wu2021self} for self-supervised contrastive learning, and SGCL~\cite{zhang2025sgcl} for supervised contrastive learning. For LLM-based baselines, we evaluate MoRec~\cite{yuan2023go}, LLMRank~\cite{hou2024large}, LLMRec~\cite{wei2024llmrec}, and TIGER~\cite{rajput2023recommender}.
To ensure fair comparison, we fix the embedding dimension to $K=128$ across all models and use MPNet~\cite{song2020mpnet} as the default embedding generator.

\begin{table*}[h]
% \small
\begin{center}
\caption{Comparisons of computation cost and performance with LLM-based RecSys in full-ranking settings.}
  \label{tab: llmrec}
  \begin{tabular}{ccccccc}
    \toprule
    \multirow{2}{*}{Methods} & \multicolumn{3}{c}{Beauty} & \multicolumn{3}{c}{Toys-Games} \\

    & {R@10} & {N@10} & \# Para.  & {R@10} & {N@10}  & \# Para.\\
    \midrule
    % TALLRec~\cite{bao2023tallrec} &  -- & -- & 7B & -- & -- & 7B \\

    MoRec~\cite{yuan2023go} &  0.0857 & 0.0397 & 13M & 0.0793 & 0.0362 & 14M \\
    LLMRank~\cite{hou2024large} & -- & -- & 7B & -- & -- & 7B\\
    LLMRec~\cite{wei2024llmrec} & 0.0974 & 0.0472 & 7B+13M & 0.0820 & 0.0389 & 7B+14M \\
    TIGER~\cite{rajput2023recommender} & 0.1015	& 0.0481	& 13M	& 0.0864 &	0.0385	& 13M \\
    \midrule
    LLMInit-Full & 0.1029 & 0.0475 & 13M & 0.0871 & 0.0387 & 14M \\
    LLMInit-Var &  0.1104 & 0.0522 & 2M & 0.0941 & 0.0421 & 2M\\
  \bottomrule
\end{tabular}
\end{center}
\end{table*}

\subsection{Performance Comparison}

The experimental results clearly demonstrate that all variants of LLMInit substantially improve recommendation performance across diverse datasets compared to traditional CF baselines. Among the proposed strategies, LLMInit-Var consistently achieves the best results, effectively capturing both collaborative filtering signals and semantic information through variance-based index selection. While both LLMInit-Rand and LLMInit-Uni offer notable gains, the variance-based method proves to be the most robust and adaptable across model types and domains.
Notably, on the Office-Products dataset (with severe embedding collapse issue as in Figuire~\ref{fig: emb_size}), LLMInit-Var delivers more than a 20\% average gain in NDCG@10 over strong baselines like SGL and SGCL. These consistent improvements highlight the generalizability and practicality of LLMInit, especially in handling diverse real-world scenarios.

\subsection{Efficiency and Perforamance Comparisons of LLM-based RecSys}
% \vspace{-10pt}

Table~\ref{tab: llmrec} compares computation cost and performance of various LLM-based RecSys in full-ranking settings. For the LLMRank~\cite{hou2024large} and LLMRec~\cite{wei2024llmrec}, we adopt the LLaMa-7B
% ~\footnote{https://huggingface.co/meta-llama/Llama-2-7b} 
model for text generation. 
LLMRank~\cite{hou2024large} directly employs an LLM for prompting but struggles with an incomplete long-context candidate pool and hallucination problem, leading to unreported performance results. 
LLMRec prompts LLM for graph augmentation, achieving recall 0.0974 (Beauty) and 0.082 (Toys) with high computational cost (7B+ parameters). 
In comparison, MoRec\cite{yuan2023go}, a modality-based model, leverages LLMs to generate textual embeddings but restricted by the use of a unified transform layer for distinct item embedding mapping. In addition, the parameter size for LLM-generated embeddings, initially around 13M/14M for MoRec and LLMInit-Full, can scale to billions as the total number of users and items surpasses 1 million, posing significant computational and storage challenges. TIGER, employing vector quantization for generative recommendation, shows potential for scalable RecSys.
LLMInit-Var, a lightweight alternative, achieves the best performance with only 2M parameters, highlighting the potential of parameter-efficient strategies in LLM-based RecSys.

\begin{figure*}[!h] 
    \centering
    \includegraphics[width=0.9\linewidth]{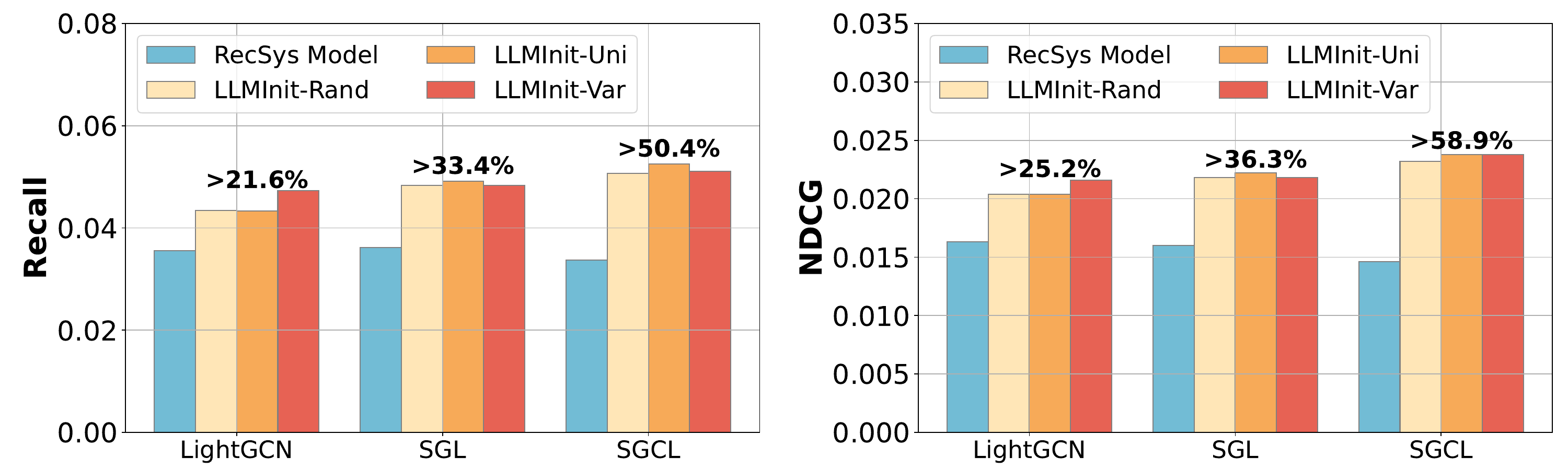}
    \caption{Performance improvement comparison in non-strict cold-start scenarios on Amazon-Beauty, where half of the observed interactions for each user are removed.
    }
\label{fig: csr}
\end{figure*}

\begin{table*}[!h]
% \small
\centering
\begin{threeparttable}
\caption{The Performance of LightGCN on Amazon-Beauty with initialization of different LLMs. The embedding size of LLMs scale from 768 to 8,096 while LLMInit models all adopt the embedding size of 128, efficient for large user base scenarios.}
\label{tab: imapct_LLM}
\begin{tabular}{c|c|ccccc}
    \toprule
    \multicolumn{2}{c}{LLMs} & MPNet\footnote{2} & gte-Qwen\footnote{3} & GPT-S\footnote{4} & GPT-L\footnote{5} & Stella\footnote{6} \\
    \midrule
    \multicolumn{2}{c}{Embedding Size} & 768 & 1,536 & 1,536 & 3,072 & 8,096\\
    \midrule
    \multirow{2}{*}{LLMInit-Rand} & Recall@10 & 0.0960 & 0.0936 &	0.0967 & 0.0925 & 0.0996\\
                        & NDCG@10 & 0.0467 & 0.0418 &	0.0463 & 0.044 & 0.0447\\
    \midrule
    \multirow{2}{*}{LLMInit-Uni} & Recall@10 & 0.1006 & 0.096 &	0.098 & 0.0942 & 0.0949\\
                        & NDCG@10 & 0.0469 & 0.0428 &	0.0465 & 0.0445 & 0.0417\\
    \midrule
    \multirow{2}{*}{LLMInit-Var} & Recall@10 & 0.1019 & 0.0932 &	0.098 & 0.096 & 0.0935\\
                        & NDCG@10 & 0.0485 & 0.0429 &	0.0465 & 0.046 & 0.0412\\
  \bottomrule
\end{tabular}
\begin{tablenotes}[flushleft] % Left-align the footnotes
    \setlength{\itemsep}{0pt} % Reduce the gap between footnotes
\footnotesize
\item[2] \url{https://huggingface.co/sentence-transformers/all-mpnet-base-v2}
\item[3] \url{https://huggingface.co/Alibaba-NLP/gte-Qwen2-1.5B-instruct}
\item[4,5] \url{https://platform.openai.com/docs/guides/embeddings/embedding-models}
\item[6] \url{https://huggingface.co/blevlabs/stella_en_v5}
\end{tablenotes}
\end{threeparttable}
\end{table*}

\subsection{Cold-Start Recommendation}

In the long-tail cold-start setting \cite{zhang2025cold, yang2025cold}, we remove half of the observed interactions of each user in training, resulting in many cold-start users with only a single interaction. Figure~\ref{fig: csr} highlights how the three CF models (LightGCN, SGL, and SGCL) react to the limited interactions and benefit from LLM-based initialization strategies in cold-start scenarios. 
LightGCN, as a simple graph model, is most resistant to the cold-start challenge, while the SGL and SGCL both suffer sharp performance drops due to reliance on interactions for data augmentation or supervised loss optimization.
With the LLMInit, SGL and SGCL demonstrate stronger abilities to utilize enhanced embeddings. In particular, SGCL achieves the highest gains with LLMInit strategies, with Recall and NDCG improving by over 50.4\% and 58.9\%, respectively. 
The supervised loss in SGCL amplifies the impact of informative embeddings, allowing it to extract more meaningful patterns from the selected dimensions. 
Overall, the results reveal that more advanced models like SGCL with nuanced supervised graph contrastive loss are better equipped to exploit high-quality LLM-based initializations.

\subsection{The impacts of Large Language Models}

Table~\ref{tab: imapct_LLM} presents the performance of LightGCN initialized with embeddings from various LLMs using different strategies on the Amazon-Beauty dataset. The results show that all tested LLMs can enhance LightGCN’s performance when paired with appropriate initialization methods. Among them, the variance-based strategy consistently delivers superior results by selecting high-variance dimensions that retain the most informative semantic features.
While larger models such as GPT-L and Stella provide higher-dimensional embeddings, their advantages are not linearly correlated with model size. Instead, MPNet and GPT-S exhibit superior performance, suggesting that the quality and domain alignment of embeddings play a more decisive role than sheer model capacity.
Consequently, optimizing initialization and embedding relevance emerges as a more effective approach than simply increasing embedding dimensionality or model size for large-scale recommendation scenarios.

\section{Conclusion}
In this work, we introduced LLMInit, a practical and scalable framework designed for industrial LLM RecSys, which effectively integrates LLM embeddings for CF model initialization. By employing selective strategies for item embeddings and aggregation pooling for user embeddings, LLMInit addresses critical industry challenges of existing CF models, such as data sparsity and cold-start scenarios, while maintaining computational efficiency suitable for large-scale deployments. Extensive evaluations on real-world datasets validate that LLMInit significantly boosts the performance of various state-of-the-art CF models, demonstrating the tangible value and feasibility of leveraging language-based world knowledge to improve recommendation quality in industrial applications.

\section*{Acknowledgment}
This work is supported in part by NSF under grants III-2106758, and POSE-2346158.

\newpage

\section*{Limitations}

Although LLMInit achieves notable improvements for scalable LLM-based recommendation, especially under cold-start and sparse-data scenarios, its effectiveness relies on the quality and domain relevance of pretrained LLM embeddings, necessitating careful selection of LLMs for domains or conducting further domain adaptation. Additionally, our current approach leverages only textual metadata; incorporating multi-modal features (such as images or audio) remains an important future direction to enhance real-world applicability. Finally, more advanced CF models such ordinary-differential-equations (ODEs)~\cite{xu2023graph, xu2025graph, zhang2024we} have not yet been evaluated within LLMInit.

\bibliography{custom}

@inproceedings{hou2024large,
  title={Large language models are zero-shot rankers for recommender systems},
  author={Hou, Yupeng and Zhang, Junjie and Lin, Zihan and Lu, Hongyu and Xie, Ruobing and McAuley, Julian and Zhao, Wayne Xin},
  booktitle={European Conference on Information Retrieval},
  pages={364--381},
  year={2024}
}

@inproceedings{bao2023tallrec,
  title={Tallrec: An effective and efficient tuning framework to align large language model with recommendation},
  author={Bao, Keqin and Zhang, Jizhi and Zhang, Yang and Wang, Wenjie and Feng, Fuli and He, Xiangnan},
  booktitle={Proceedings of the 17th ACM Conference on Recommender Systems},
  pages={1007--1014},
  year={2023}
}

@article{zhang2025personaagent,
  title={Personaagent: When large language model agents meet personalization at test time},
  author={Zhang, Weizhi and Zhang, Xinyang and Zhang, Chenwei and Yang, Liangwei and Shang, Jingbo and Wei, Zhepei and Zou, Henry Peng and Huang, Zijie and Wang, Zhengyang and Gao, Yifan and others},
  journal={arXiv preprint arXiv:2506.06254},
  year={2025}
}

@article{yang2025cold,
  title={Cold-Start Recommendation with Knowledge-Guided Retrieval-Augmented Generation},
  author={Yang, Wooseong and Zhang, Weizhi and Liu, Yuqing and Han, Yuwei and Wang, Yu and Lee, Junhyun and Yu, Philip S},
  journal={arXiv preprint arXiv:2505.20773},
  year={2025}
}

@article{yang2024item,
  title={Item Cluster-aware Prompt Learning for Session-based Recommendation},
  author={Yang, Wooseong and Wang, Chen and Song, Zihe and Zhang, Weizhi and Yu, Philip S},
  journal={arXiv preprint arXiv:2410.04756},
  year={2024}
}

@article{liu2024towards,
  title={Towards Neural Scaling Laws on Graphs},
  author={Liu, Jingzhe and Mao, Haitao and Chen, Zhikai and Zhao, Tong and Shah, Neil and Tang, Jiliang},
  journal={arXiv preprint arXiv:2402.02054},
  year={2024}
}

@inproceedings{zhang2024we,
  title={Do We Really Need Graph Convolution During Training? Light Post-Training Graph-ODE for Efficient Recommendation},
  author={Zhang, Weizhi and Yang, Liangwei and Song, Zihe and Zou, Henry Peng and Xu, Ke and Fang, Liancheng and Yu, Philip S},
  booktitle={Proceedings of the 33rd ACM International Conference on Information and Knowledge Management},
  pages={3248--3258},
  year={2024}
}

@article{song2020mpnet,
  title={Mpnet: Masked and permuted pre-training for language understanding},
  author={Song, Kaitao and Tan, Xu and Qin, Tao and Lu, Jianfeng and Liu, Tie-Yan},
  journal={Advances in neural information processing systems},
  volume={33},
  pages={16857--16867},
  year={2020}
}

@article{sheng2024language,
  title={Language Representations Can be What Recommenders Need: Findings and Potentials},
  author={Sheng, Leheng and Zhang, An and Zhang, Yi and Chen, Yuxin and Wang, Xiang and Chua, Tat-Seng},
  journal={arXiv preprint arXiv:2407.05441},
  year={2024}
}

@article{koren2021advances,
  title={Advances in collaborative filtering},
  author={Koren, Yehuda and Rendle, Steffen and Bell, Robert},
  journal={Recommender systems handbook},
  pages={91--142},
  year={2021},
  publisher={Springer}
}

@inproceedings{zhang2023dual,
  title={Dual-Teacher Knowledge Distillation for Strict Cold-Start Recommendation},
  author={Zhang, Weizhi and Yang, Liangwei and Cao, Yuwei and Xu, Ke and Zhu, Yuanjie and Philip, S Yu},
  booktitle={2023 IEEE International Conference on Big Data (BigData)},
  pages={483--492},
  year={2023},
  organization={IEEE}
}

@article{zhang2024mixed,
  title={Mixed Supervised Graph Contrastive Learning for Recommendation},
  author={Zhang, Weizhi and Yang, Liangwei and Song, Zihe and Zou, Henry Peng and Xu, Ke and Zhu, Yuanjie and Yu, Philip S},
  journal={arXiv preprint arXiv:2404.15954},
  year={2024}
}

@article{kaplan2020scaling,
  title={Scaling laws for neural language models},
  author={Kaplan, Jared and McCandlish, Sam and Henighan, Tom and Brown, Tom B and Chess, Benjamin and Child, Rewon and Gray, Scott and Radford, Alec and Wu, Jeffrey and Amodei, Dario},
  journal={arXiv preprint arXiv:2001.08361},
  year={2020}
}

@inproceedings{wang2021dcn,
  title={Dcn v2: Improved deep \& cross network and practical lessons for web-scale learning to rank systems},
  author={Wang, Ruoxi and Shivanna, Rakesh and Cheng, Derek and Jain, Sagar and Lin, Dong and Hong, Lichan and Chi, Ed},
  booktitle={Proceedings of the web conference 2021},
  pages={1785--1797},
  year={2021}
}

@inproceedings{wei2024llmrec,
  title={Llmrec: Large language models with graph augmentation for recommendation},
  author={Wei, Wei and Ren, Xubin and Tang, Jiabin and Wang, Qinyong and Su, Lixin and Cheng, Suqi and Wang, Junfeng and Yin, Dawei and Huang, Chao},
  booktitle={Proceedings of the 17th ACM International Conference on Web Search and Data Mining},
  pages={806--815},
  year={2024}
}

@inproceedings{liang-etal-2025-taxonomy,
    title = "Taxonomy-Guided Zero-Shot Recommendations with {LLM}s",
    author = "Liang, Yueqing  and
      Yang, Liangwei  and
      Wang, Chen  and
      Xu, Xiongxiao  and
      Yu, Philip S.  and
      Shu, Kai",
    booktitle = "Proceedings of the 31st International Conference on Computational Linguistics",
    month = jan,
    year = "2025",
    address = "Abu Dhabi, UAE",
    publisher = "Association for Computational Linguistics",
    url = "https://aclanthology.org/2025.coling-main.102/",
    pages = "1520--1530"
}

@inproceedings{xu2023graph,
  title={Graph Neural Ordinary Differential Equations-based method for Collaborative Filtering},
  author={Xu, Ke and Zhu, Yuanjie and Zhang, Weizhi and Philip, S Yu},
  booktitle={2023 IEEE International Conference on Data Mining (ICDM)},
  pages={1445--1450},
  year={2023},
  organization={IEEE}
}

@inproceedings{xu2025graph,
  title={Graph Neural Controlled Differential Equations For Collaborative Filtering},
  author={Xu, Ke and Zhang, Weizhi and Song, Zihe and Zhu, Yuanjie and Yu, Philip S},
  booktitle={Companion Proceedings of the ACM on Web Conference 2025},
  pages={1446--1449},
  year={2025}
}

@inproceedings{wang2019neural,
  title={Neural graph collaborative filtering},
  author={Wang, Xiang and He, Xiangnan and Wang, Meng and Feng, Fuli and Chua, Tat-Seng},
  booktitle={Proceedings of the 42nd international ACM SIGIR conference on Research and development in Information Retrieval},
  pages={165--174},
  year={2019}
}

@inproceedings{he2020lightgcn,
  title={Lightgcn: Simplifying and powering graph convolution network for recommendation},
  author={He, Xiangnan and Deng, Kuan and Wang, Xiang and Li, Yan and Zhang, Yongdong and Wang, Meng},
  booktitle={Proceedings of the 43rd International ACM SIGIR conference on research and development in Information Retrieval},
  pages={639--648},
  year={2020}
}

@article{rendle2012bpr,
  title={BPR: Bayesian personalized ranking from implicit feedback},
  author={Rendle, Steffen and Freudenthaler, Christoph and Gantner, Zeno and Schmidt-Thieme, Lars},
  journal={arXiv preprint arXiv:1205.2618},
  year={2012}
}

@inproceedings{fan2019graph,
  title={Graph neural networks for social recommendation},
  author={Fan, Wenqi and Ma, Yao and Li, Qing and He, Yuan and Zhao, Eric and Tang, Jiliang and Yin, Dawei},
  booktitle={The world wide web conference},
  pages={417--426},
  year={2019}
}

@inproceedings{wu2021self,
  title={Self-supervised graph learning for recommendation},
  author={Wu, Jiancan and Wang, Xiang and Feng, Fuli and He, Xiangnan and Chen, Liang and Lian, Jianxun and Xie, Xing},
  booktitle={Proceedings of the 44th international ACM SIGIR conference on research and development in information retrieval},
  pages={726--735},
  year={2021}
}

@article{zhang2025cold,
  title={Cold-Start Recommendation towards the Era of Large Language Models (LLMs): A Comprehensive Survey and Roadmap},
  author={Zhang, Weizhi and Bei, Yuanchen and Yang, Liangwei and Zou, Henry Peng and Zhou, Peilin and Liu, Aiwei and Li, Yinghui and Chen, Hao and Wang, Jianling and Wang, Yu and others},
  journal={arXiv preprint arXiv:2501.01945},
  year={2025}
}

@article{achiam2023gpt,
  title={Gpt-4 technical report},
  author={Achiam, Josh and Adler, Steven and Agarwal, Sandhini and Ahmad, Lama and Akkaya, Ilge and Aleman, Florencia Leoni and Almeida, Diogo and Altenschmidt, Janko and Altman, Sam and others},
  journal={arXiv preprint arXiv:2303.08774},
  year={2023}
}

@article{touvron2023llama,
  title={Llama: Open and efficient foundation language models},
  author={Touvron, Hugo and Lavril, Thibaut and Izacard, Gautier and Martinet, Xavier and Lachaux, Marie-Anne and Lacroix, Timoth{\'e}e and Rozi{\`e}re, Baptiste and Goyal, Naman and Hambro, Eric and Azhar, Faisal and others},
  journal={arXiv preprint arXiv:2302.13971},
  year={2023}
}

@inproceedings{geng2022recommendation,
  title={Recommendation as language processing (rlp): A unified pretrain, personalized prompt \& predict paradigm (p5)},
  author={Geng, Shijie and Liu, Shuchang and Fu, Zuohui and Ge, Yingqiang and Zhang, Yongfeng},
  booktitle={Proceedings of the 16th ACM Conference on Recommender Systems},
  pages={299--315},
  year={2022}
}

@inproceedings{li2023text,
  title={Text is all you need: Learning language representations for sequential recommendation},
  author={Li, Jiacheng and Wang, Ming and Li, Jin and Fu, Jinmiao and Shen, Xin and Shang, Jingbo and McAuley, Julian},
  booktitle={Proceedings of the 29th ACM SIGKDD Conference on Knowledge Discovery and Data Mining},
  pages={1258--1267},
  year={2023}
}

@inproceedings{yuan2023go,
  title={Where to go next for recommender systems? id-vs. modality-based recommender models revisited},
  author={Yuan, Zheng and Yuan, Fajie and Song, Yu and Li, Youhua and Fu, Junchen and Yang, Fei and Pan, Yunzhu and Ni, Yongxin},
  booktitle={Proceedings of the 46th International ACM SIGIR Conference on Research and Development in Information Retrieval},
  pages={2639--2649},
  year={2023}
}

@inproceedings{guo2024embedding,
  title={On the Embedding Collapse when Scaling up Recommendation Models},
  author={Guo, Xingzhuo and Pan, Junwei and Wang, Ximei and Chen, Baixu and Jiang, Jie and Long, Mingsheng},
  booktitle={Forty-first International Conference on Machine Learning}
}

@article{wu2024survey,
  title={A survey on large language models for recommendation},
  author={Wu, Likang and Zheng, Zhi and Qiu, Zhaopeng and Wang, Hao and Gu, Hongchao and Shen, Tingjia and Qin, Chuan and Zhu, Chen and Zhu, Hengshu and Liu, Qi and others},
  journal={World Wide Web},
  volume={27},
  number={5},
  pages={60},
  year={2024},
  publisher={Springer}
}

@inproceedings{hou2022towards,
  title={Towards universal sequence representation learning for recommender systems},
  author={Hou, Yupeng and Mu, Shanlei and Zhao, Wayne Xin and Li, Yaliang and Ding, Bolin and Wen, Ji-Rong},
  booktitle={Proceedings of the 28th ACM SIGKDD Conference on Knowledge Discovery and Data Mining},
  pages={585--593},
  year={2022}
}

@inproceedings{muennighoff2023mteb,
  title={MTEB: Massive Text Embedding Benchmark},
  author={Muennighoff, Niklas and Tazi, Nouamane and Magne, Loic and Reimers, Nils},
  booktitle={Proceedings of the 17th Conference of the European Chapter of the Association for Computational Linguistics},
  pages={2014--2037},
  year={2023}
}

@inproceedings{ni2019justifying,
  title={Justifying recommendations using distantly-labeled reviews and fine-grained aspects},
  author={Ni, Jianmo and Li, Jiacheng and McAuley, Julian},
  booktitle={Proceedings of the 2019 conference on empirical methods in natural language processing and the 9th international joint conference on natural language processing (EMNLP-IJCNLP)},
  pages={188--197},
  year={2019}
}

@article{rajput2023recommender,
  title={Recommender systems with generative retrieval},
  author={Rajput, Shashank and Mehta, Nikhil and Singh, Anima and Hulikal Keshavan, Raghunandan and Vu, Trung and Heldt, Lukasz and Hong, Lichan and Tay, Yi and Tran, Vinh and Samost, Jonah and others},
  journal={Advances in Neural Information Processing Systems},
  volume={36},
  pages={10299--10315},
  year={2023}
}

@inproceedings{zhang2025sgcl,
  title={SGCL: Unifying Self-Supervised and Supervised Learning for Graph Recommendation},
  author={Zhang, Weizhi and Yang, Liangwei and Song, Zihe and Zou, Henry Peng and Xu, Ke and Zhu, Yuanjie and Yu, Philip S},
  booktitle={Proceedings of the Nineteenth ACM Conference on Recommender Systems},
  pages={671--676},
  year={2025}
}

@inproceedings{qiu2021u,
  title={U-BERT: Pre-training user representations for improved recommendation},
  author={Qiu, Zhaopeng and Wu, Xian and Gao, Jingyue and Fan, Wei},
  booktitle={Proceedings of the AAAI Conference on Artificial Intelligence},
  volume={35},
  number={5},
  pages={4320--4327},
  year={2021}
}

@inproceedings{xi2024towards,
  title={Towards open-world recommendation with knowledge augmentation from large language models},
  author={Xi, Yunjia and Liu, Weiwen and Lin, Jianghao and Cai, Xiaoling and Zhu, Hong and Zhu, Jieming and Chen, Bo and Tang, Ruiming and Zhang, Weinan and Yu, Yong},
  booktitle={Proceedings of the 18th ACM Conference on Recommender Systems},
  pages={12--22},
  year={2024}
}

\appendix

\section{Dataset Statistics}
\label{sec:app_data}
\begin{table}[htpb]
  \centering
  \caption{Statistics of the datasets.}
  \label{tab:data}
  \begin{adjustbox}{max width=\linewidth}
    \begin{tabular}{lcccc}
      \toprule
      Dataset & \# Users & \# Items & \# Inter. & Sparsity \\
      \midrule
      Beauty           & 10,554 & 6,087 & 94,148 & 99.85\% \\
      Toys–Games       & 11,269 & 7,310 & 95,420 & 99.88\% \\
      Tools–Home       &  9,246 & 6,199 & 73,250 & 99.87\% \\
      Office–Products  &  3,261 & 1,584 & 36,544 & 99.29\% \\
      \bottomrule
    \end{tabular}
  \end{adjustbox}
\end{table}

To comprehensively evaluate the effectiveness of LLM-based initialization strategies in real-world recommendation scenarios, we conduct experiments on four widely used Amazon product review datasets~\citep{ni2019justifying}, covering diverse user interaction domains. Each dataset consists of implicit user–item interactions, where an interaction denotes a user’s engagement (e.g., rating and review) with a product. Following common practice, we filter out users and items with fewer than five interactions and adopt the leave-one-out evaluation protocol. The statistics of the processed datasets are summarized in Table~\ref{tab:data}.

\section{Pooling Strategies}
\label{sec:app_data}
\begin{table}[t]
\centering
\caption{Performance comparison of different initialization pooling strategies on Amazon-Beauty. All models are initialized with LLM-based embeddings except for the random baseline.}
\label{tab:pooling_strategies}
\renewcommand{\arraystretch}{1.3}
\begin{tabular}{lcc}
\toprule
\textbf{Method} & \textbf{Recall@10} & \textbf{NDCG@10} \\
\midrule
Random Init & 0.1029 & 0.0475 \\
Mean (LLMInit) & \textbf{0.1104} & \textbf{0.0522} \\
Max (LLMInit) & 0.1088 & 0.0510 \\
LightGCN (LLMInit) & 0.1101 & 0.0515 \\
\bottomrule
\end{tabular}
\end{table}

To further analyze the impact of different embedding aggregation strategies, Table~\ref{tab:pooling_strategies} compares pooling-based methods for user embedding initializations. Results show that all LLM-based initialization strategies outperform random initialization, confirming the benefits of semantic priors learned by LLMs. Mean pooling achieves the best overall performance, slightly surpassing both max pooling and LightGCN pooling. This indicates that averaging item embeddings provides a stable and representative initialization signal for graph-based user–item interaction modeling. 

% This is an appendix.

\end{document}